\documentclass[aps,twocolumn,showpacs,superscriptaddress]{revtex4}
\usepackage{graphicx,color}
\usepackage{amsmath}
\usepackage{amssymb}

\usepackage{amsfonts}
\usepackage{bm}

\newcommand{\ot}{{\,\otimes\,}}
\newcommand{{\Cd}}{{\mathbb{C}^d}}

\def\oper{{\mathchoice{\rm 1\mskip-4mu l}{\rm 1\mskip-4mu l}%
{\rm 1\mskip-4.5mu l}{\rm 1\mskip-5mu l}}}
\def\<{\langle}
\def\>{\rangle}

\newtheorem{example}{Example}

\begin{document}
\title{Non-Markovian quantum dynamics: local versus non-local }
\author{Dariusz Chru\'sci\'nski and Andrzej Kossakowski}
\affiliation{Institute of Physics, Nicolaus Copernicus University \\
Grudzi\c{a}dzka 5/7, 87--100 Toru\'n, Poland}

\begin{abstract}
We analyze non-Markovian evolution of open quantum systems. It is
shown that any dynamical map representing evolution of such a system
may be described either by non-local  master equation with memory
kernel or equivalently by equation which is local in time. These two
descriptions are complementary: if one is simple the other is quite
involved, or even singular, and vice versa. The price one pays for
the local approach is that the corresponding generator keeps the
memory about the starting point `$t_0$'. This is the very essence of
non-Markovianity. Interestingly, this generator might be highly
singular, nevertheless,  the corresponding dynamics is perfectly
regular. Remarkably, singularities of generator may lead to
interesting physical phenomena like revival of coherence or sudden
death and revival of entanglement.
\end{abstract}


\maketitle

The non-Markovian dynamics of open quantum systems attracts nowadays
increasing attention \cite{Breuer}. It is very much connected to the
growing interest in controlling quantum systems and applications in
modern quantum technologies such as quantum communication,
cryptography and computation \cite{QIT}. It turns out that the
popular Markovian approximation which does not take into account
memory effects is not sufficient for modern applications and todays
technology calls for truly non-Markovian approach. Non-Markovian
dynamics was recently studied in
\cite{Wilkie,Budini,B-2004,Wodkiewicz,Lidar,Maniscalco1,Maniscalco2,Maniscalco-09,KR,KR-last,B,Francesco,DAS}.
Interestingly, several measures of non-Markovianity were proposed
during last year \cite{Wolf,Breuer-09,Plenio,Chiny-09}.

The standard approach to the dynamics of open system uses the
Nakajima-Zwanzig projection operator technique \cite{NZ} which shows
that under fairly general conditions, the master equation for the
reduced density matrix $\rho(t)$ takes the form of the following
non-local equation
\begin{equation}\label{NZ}
\frac{d}{dt}\, \rho(t) = \int_{t_0}^t \mathcal{K}(t - u)\rho(u)\,
du\ , \ \ \ \rho(t_0)=\rho_0\ ,
\end{equation}
in which quantum memory effects are taken into account through the
introduction of the memory kernel $\mathcal{K}(t)$: this simply
means that the rate of change of the state $\rho(t)$ at time $t$
depends on its history (starting at $t=t_0$). Usually, one takes
$t_0=0$, however, in this letter we shall keep `$t_0$' arbitrary. An
alternative and technically much simpler scheme is the
time-convolutionless (TCL) projection operator technique
\cite{TCL,BKP,Breuer} in which one obtains a first-order
differential equation for the reduced density matrix. The advantage
of the TCL approach consists in the fact that it yields an equation
of motion for the relevant degrees of freedom which is local in time
and which is therefore often much easier to deal with than the
Nakajima-Zwanzig non-local master equation (\ref{NZ}).

An essential step to derive TCL from (\ref{NZ}) relies on the
existence of certain operator inverse \cite{BKP}. However, this
inverse needs not exist and then the method does not work
\cite{Breuer,BKP}. Moreover, even if it exists the corresponding
local in time TCL generator is usually defined by the perturbation
series (see e.g. detailed discussion in \cite{Breuer}) in powers of
the coupling strength characterizing the system. However in general
the perturbative approach  leads to significant problems. For
example the dynamical map needs not be completely positive if one
takes only finite number of terms from the perturbative expansion.

In the present paper we take a different path. We show that any
solution of the non-local equation (\ref{NZ}) always satisfies
equation which does not involve the integral memory kernel, i.e. it
is local in time. However, the corresponding generator is
effectively non-local due to the fact that it keeps the memory about
the starting point $t_0$. Moreover, as we shall see, this generator
may be singular, nevertheless, it always leads to perfectly regular
dynamics.

Let us start with the standard Markovian master equation
\begin{equation}\label{M}
    \frac{d\rho(t)}{dt} = \mathcal{L}\, \rho(t)  \ , \ \ \ \  \rho(t_0)=\rho_0\
    ,
\end{equation}
where $\cal L$ is a time-independent generator possessing the
following well known representation \cite{Lindblad,Alicki}
\begin{equation}\label{GENERATOR}
    \mathcal{L}\rho = -i [H,\rho] + \sum_\alpha \left( V_\alpha \rho
    V_\alpha^\dagger - \frac 12 \{ V^\dagger_\alpha V_\alpha,\rho\} \right) \
    .
\end{equation}
The above structure of $\cal L$ guaranties that dynamical map
$\Lambda(t,t_0)$, defined by $\rho(t) = \Lambda(t,t_0)\rho_0$, is
completely positive and trace preserving for $t\geq t_0$. Note that
$\Lambda(t,t_0)$ itself satisfies Markovian master equation
\begin{equation}\label{D-M}
    \frac{d}{dt}\Lambda(t,t_0) = {\cal L}\, \Lambda(t,t_0)  \ , \ \ \ \  \Lambda(t_0,t_0)=\oper\
    ,
\end{equation}
and the solution for $\Lambda(t,t_0)$ is given by
    $\Lambda(t,t_0) = e^{(t-t_0){\cal L}}$,
which implies that $\Lambda(t,t_0)$ depends only upon the difference
`$t-t_0$' and hence $\Lambda(t) := \Lambda(t,0)$ defines a
1-parameter semigroup satisfying homogeneous composition law
\begin{equation}\label{COMP}
    \Lambda(t_1) \Lambda(t_2) = \Lambda(t_1+t_2) \ ,
\end{equation}
for $t_1,t_2\geq 0$. In general the external conditions which
influence the dynamics of an open system may very in time. The
natural generalization of the Markovian master equation (\ref{M})
involves time-dependent generator $\mathcal{L}(t)$ which has exactly
the same representation as in (\ref{GENERATOR}) with time-dependent
Hamiltonian $H(t)$ and time-dependent Lindblad operators
$V_\alpha(t)$. Therefore one gets the following master equation for
the dynamical map $\Lambda(t,t_0)$
\begin{equation}\label{D-M-t}
    \frac{d}{dt}\Lambda(t,t_0) = {\cal L}(t)\, \Lambda(t,t_0)  \ , \ \ \ \  \Lambda(t_0,t_0)=\oper\
    ,
\end{equation}
with leads to the following solution
\begin{equation}\label{T-local}
    \Lambda(t,t_0) = {\rm T}\, \exp\left(\int_{t_0}^t{\cal L}(\tau)d\tau\right) ,
\end{equation}
where T stands for the chronological operator. Clearly,
$\Lambda(t,t_0)$ no longer depends upon `$t-t_0$' but it still
satisfies inhomogeneous composition law
\begin{equation}\label{COMP-t}
    \Lambda(t,s) \cdot \Lambda(s,t_0) = \Lambda(t,t_0) \ ,
\end{equation}
for $t\geq s\geq t_0$. We stress that (\ref{D-M-t}) although
time-dependent is perfectly Markovian.


Let us turn to the non-Markovian evolution (\ref{NZ}). One obtains
the following equation for the corresponding dynamical map
\begin{equation}\label{D-NM}
    \frac{d}{dt}\Lambda(t,t_0) = \int_{t_0}^t d\tau\, \mathcal{K}(t - \tau) \,
    \Lambda(\tau,t_0)  \ , \ \   \Lambda(t_0,t_0)=\oper\ .
\end{equation}
Now comes an essential observation: $\Lambda(t,t_0)$ does depend
upon the difference `$t-t_0$' and hence it shares the same feature
as the Markovian dynamics with time independent generator (\ref{M}).
The proof is very easy. Observe that any non-Markovian dynamics in
$\mathcal{H}$ may be defined as a reduced Markovian dynamics on the
extended Hilbert space $\mathcal{H} \ot \mathcal{H}_a$
($\mathcal{H}_a$ denotes ancilla Hilbert space). If $\omega$ denotes
a fixed state of the ancilla, then
\begin{equation}\label{reduced}
    \Lambda(t,t_0)\rho := {\rm Tr}_a [ e^{(t-t_0)L} (\rho \ot \omega)
    ]\ ,
\end{equation}
where we trace out over ancilla degrees of freedom and $L$ denotes
the total Markovian generator in $\mathcal{H} \ot \mathcal{H}_a$.
Since the r.h.s of (\ref{reduced}) depends on `$t-t_0$' so does the
non-Markovian dynamical map  $\Lambda(t,t_0)$.  Hence, the
non-Markovian dynamics is homogeneous (depends on $t-t_0$) but of
course does not satisfy the composition law (\ref{COMP}). This is
the very essence of non-Markovianity and it does provide the evident
sign of the memory.

Suppose now that $\Lambda(t,t_0)$ satisfies non-local equation
(\ref{D-NM}). Taking into account that $\Lambda$ is a function of
$\tau=t-t_0$,  let us consider its spectral decomposition
\begin{equation}\label{}
\Lambda(\tau)\rho = \sum_\mu \lambda_\mu(\tau)\, F_\mu(\tau) {\rm
Tr}(G_\mu^\dagger(\tau) \rho)\ ,
\end{equation}
where $F_\mu(\tau)$ and $G_\mu(\tau)$ define the damping basis for
$\Lambda(\tau)$, that is, ${\rm Tr}(F_\mu(\tau) G_\nu^\dagger(\tau))
= \delta_{\mu\nu}$. Clearly, for $\tau=0$ one has $\lambda_\mu(0) =
1$. Now one defines the formal inverse
\begin{equation}\label{}
\Lambda^{-1}(\tau)\rho = \sum_\mu \lambda^{-1}_\mu(\tau)\,
F_\mu(\tau) {\rm Tr}(G_\mu^\dagger(\tau) \rho)\ ,
\end{equation}
such that $\Lambda(\tau) \Lambda^{-1}(\tau) = \oper$ for $\tau \geq
0$.
It should be stressed that $\Lambda^{-1}(\tau)$ needs not exist (it
does exist if and only if $\lambda_\mu(\tau) \neq 0$). Moreover, the
existence of $\Lambda^{-1}(\tau)$ does not mean that the dynamics is
invertible. Note, that even if $\Lambda^{-1}(\tau)$ does exist it is
in general not completely positive and hence can not describe
quantum evolution backwards in time. Actually, $\Lambda^{-1}(\tau)$
is completely positive if and only if $\Lambda(\tau)$ is unitary or
anti-unitary. In  this case $|\lambda_\mu(\tau)|=1$ and
$\lambda_\mu^{-1}(\tau) = \overline{\lambda_\mu(\tau)}$. It is
therefore clear that the non-local equation (\ref{D-NM}) reduces
formally to the following one
\begin{equation}\label{local-NM}
     \frac{d}{dt}\Lambda(t,t_0) = \mathcal{L}(t-t_0) \Lambda(t,t_0)\
     , \ \ \Lambda(t_0,t_0) = \oper\ ,
\end{equation}
where the time-dependent generator $\mathcal{L}(\tau)$ is defined by
the following logarithmic derivative of the dynamical map
\begin{equation}\label{}
    \mathcal{L}(\tau) := \frac{d}{d\tau}\Lambda(\tau) \cdot
    \Lambda^{-1}(\tau)\ .
\end{equation}
One easily finds the following formula
\begin{equation}\label{}
    \mathcal{L}(\tau)\rho = \sum_{\mu\nu} \mathcal{L}_{\mu\nu}(\tau) {\rm
    Tr}(G_\nu^\dagger(\tau) \rho) \ ,
\end{equation}
with
\begin{equation*}\label{}
    \mathcal{L}_{\mu\nu} =
    \left(\frac{\dot{\lambda}_\mu}{\lambda_\nu}F_\mu  +
    \frac{{\lambda}_\mu}{\lambda_\nu}\dot{F}_\mu\right) \delta_{\mu\nu}
    + \frac{{\lambda}_\mu}{\lambda_\nu} F_\mu {\rm
    Tr}(\dot{G}_\mu^\dagger F_\nu)\ ,
\end{equation*}
where for simplicity we omit the time dependence.  In particular, if
the damping basis is time-independent, and $\lambda_\mu(\tau) =
e^{\gamma_\mu(\tau)}$, then the spectral decomposition of
$\mathcal{L}(\tau)$ has a particulary simple form
\begin{equation}\label{}
    \mathcal{L}(\tau) \rho = \sum_{\mu} \dot{\gamma}_\mu(\tau) F_\mu  {\rm
    Tr}(G_\mu ^\dagger \rho) \ .
\end{equation}
Summarizing, we have shown that each solution $\Lambda(t,t_0)$ to
the non-local non-Markovian equation (\ref{D-NM})  does satisfy the
first order differential   equation (\ref{local-NM}). Let us observe
that Eq. (\ref{local-NM}) is local in time but its generator does
remember about the starting point `$t_0$'. This is the most
important difference with the time-dependent Markovian equation
(\ref{D-M-t}). The appearance of `$t_0$' in the generator
$\mathcal{L}(t-t_0)$ implies that $\mathcal{L}$ is effectively
non-local in time, that is,  it contains a memory. Therefore, the
local equation (\ref{local-NM}) is non-Markovian contrary to the
local equation (\ref{D-M-t}) which does does not keep any memory
about $t_0$. Note, that solution to (\ref{local-NM}) is given by
\begin{equation}\label{}
    \Lambda(t,t_0) = {\rm T}\, \exp\left( \int_{0}^{t-t_0}
    \mathcal{L}(\tau)\, d\tau\right) \ .
\end{equation}
It shows that $\Lambda(t,t_0)$ is indeed homogeneous in time
(depends on `$t-t_0$'). However, contrary to (\ref{T-local}), it
does not satisfy the composition law. Again, this is a clear sign
for the memory effect.


One may ask a natural question: how to construct non-Markovian
generator $\mathcal{L}(\tau)$. The general answer is not known but
one may easily propose special constructions. Let $\mathcal{L}$ be a
Markovian generator defined by (\ref{GENERATOR}) and define
$\mathcal{L}(\tau) = \alpha(\tau) \mathcal{L}$. It is clear that if
$\int_0^\tau \alpha(u)du \geq 0$ for $\tau \geq 0$, then
$\Lambda(\tau) = \exp( \int_0^\tau \alpha(u)du \, \mathcal{L})$
defines completely positive non-Markovian dynamics. This
construction may be generalized as follows: consider $N$ mutually
commuting Markovian generators $\mathcal{L}_1,\ldots,\mathcal{L}_N$
and $N$ real functions $\alpha_k(\tau)$ satisfying $\int_0^\tau
\alpha_k(u)du \geq 0$. Then ${\cal L}(\tau) = \alpha_1(\tau)
\mathcal{L}_1 + \ldots + \alpha_N(\tau) \mathcal{L}_N$ serves as a
generator of non-Markovian evolution. Finally, let us observe that
if $\mathcal{L}(t)$ is a time-dependent Markovian generator (i.e. it
has the Lindblad form (\ref{GENERATOR})  with time-dependent
Hamiltonian $H(t)$  and noise operators $V_\alpha(t)$), then
$\mathcal{L}(t-t_0)$ generates the non-Markovian dynamics for $t
\geq t_0$. We stress that these constructions provide only
restricted classes of examples of non-Markovian generators. All of
them start with  a set of Markovian generators and produce a
non-Markovian one. It turns out (see Example \ref{E-PURE} below)
that one may construct generators which do not fit these classes.


Let us illustrate our analysis with the following simple examples.

\begin{example}
{\em Consider the dynamical map for a qudit ($d$-level quantum
system) given by
\begin{equation}\label{EX-1}
    \Lambda(\tau) = \left(1 - \int_0^{\tau} f(u)du \right) \oper +  \int_0^{\tau}
    f(u)du\, \mathcal{P} \ ,
\end{equation}
where $\mathcal{P} : \mathcal{B}(\mathbb{C}^d) \longrightarrow
\mathcal{B}(\mathbb{C}^d)$ denotes  completely positive trace
preserving projection. For example take a fixed qudit state $\omega$
and define $\mathcal{P}$ by the following formula $\mathcal{P} \rho
= \omega\, {\rm Tr}\rho$. The real function `$f$' satisfies:
\[ 0\ \leq\ \int_0^\tau f(u)du \ \leq\  1\ , \]
for any $\tau > 0$. Note that $f(u)$ needs not be positive. If $f(u)
\geq 0$ ($u\geq 0$), then
 $\Lambda(\tau)$ defines quantum semi-Markov process
and the function $f(u)$ may be interpreted as a waiting time
distribution for this process \cite{Budini,B}. Clearly,
$\Lambda(\tau)$ being a convex combination of $\oper$ and
$\mathcal{P}$ is completely positive trace preserving map and hence
it defines legal quantum dynamics of a qudit. The corresponding
memory kernel is well known \cite{Budini,B} and it is given
\begin{equation}\label{}
    \mathcal{K}(\tau) =  \kappa(\tau) \mathcal{L}_0\, \ ,
\end{equation}
where the function $\kappa(\tau)$ is defined in terms of its Laplace
transform as follows
\begin{equation}\label{kappa-1}
    \widetilde{\kappa}(s) = \frac{s \widetilde{f}(s)}{1 -
    \widetilde{f}(s)} \ ,
\end{equation}
and $\mathcal{L}_0$ is defined by $\mathcal{L}_0 = \mathcal{P} -
\oper$.  Note, that $\mathcal{L}_0$ has exactly the structure of the
Markovian generator (\ref{GENERATOR}) with $H=0$, and the Lindblad
operators $V_\alpha$ define Kraus representation of $\mathcal{P}$,
that is $\mathcal{P}\rho = \sum_\alpha V_\alpha \rho
V_\alpha^\dagger$. One easily finds for the corresponding generator
\begin{equation}\label{}
    \mathcal{L}(\tau) =  \alpha(\tau) \, \mathcal{L}_0\ ,
\end{equation}
where
\begin{equation}\label{alpha-1}
    \alpha(\tau) = \frac{f(\tau)}{1 - \int_0^{\tau} f(u)du} \ .
\end{equation}
 Let us observe that
\begin{equation*}\label{}
    \int_0^\tau \alpha(u)du = - \ln\Big(1- \int_0^\tau f(u)du\Big) \geq 0 \
    ,
\end{equation*}
and hence this example gives rise to $ \mathcal{L}(\tau) =
\alpha(\tau) \mathcal{L}_0$, with Markovian $\mathcal{L}_0$ and
$\alpha(\tau)$ satisfying  $\int_0^\tau \alpha(u)du \geq 0$. We
stress that $\alpha(\tau)$ needs not be positive. It is positive if
and only if $f(\tau)$ corresponds to the waiting time distribution
\cite{Budini,B}. Note the striking similarity between formulae
(\ref{kappa-1}) and (\ref{alpha-1}). It should be stressed that in
this case one knows an explicit formula for time-local generator
$\mathcal{L}(\tau)$. Note, however, that in general one is not able
to invert the Laplace transform of $\widetilde{\kappa}(s)$ from the
formula (\ref{kappa-1}) and hence the explicit formula for the
memory kernel $\mathcal{K}(t)$ is not known.

 }
\end{example}
\begin{example}
{\em  The previous example may be easily generalized to bipartite
systems. Consider for example a 2-qubit system and let $\mathcal{P}$
be a projector onto the diagonal part with respect to the product
basis $|m\ot n\>$ in $\mathbb{C}^2 \ot \mathbb{C}^2$. Let us take as
an initial density matrix so called $X$-state \cite{Eberly}
represented by
\begin{equation}\label{X}
    \rho_0 = \left( \begin{array}{cccc} \rho_{11} & 0 & 0 & \rho_{14} \\
0 & \rho_{22} & \rho_{23} & 0 \\
0 & \rho_{32} & \rho_{33} & 0 \\
\rho_{41} & 0 & 0 & \rho_{44} \end{array} \right) \ .
\end{equation}
It is easy to see that $\Lambda(\tau)$ defined by (\ref{EX-1}) does
preserve the structure of $X$-state, that is, $\rho(\tau)$ has
exactly the same form as in (\ref{X}) with $\tau$-dependent
$\rho_{mn}$. It is clear that the diagonal elements are time
independent $\rho_{kk}(\tau) = \rho_{kk}$, and $\rho_{kl}(\tau) = (1
- \int_0^{\tau} f(u)du ) \rho_{kl}$, for $k \neq l$. The
entanglement of the 2-qubit $X$-state $\rho(\tau)$ is uniquely
determined by the concurrence $\, C(\tau) = 2\max\{
c_1(\tau),c_2(\tau),0\}\,$, where
\[  c_1(\tau) = |\rho_{23}(\tau)| - \sqrt{\rho_{11}\rho_{44}} \ , \ \
c_2(\tau) = |\rho_{14}(\tau)| - \sqrt{\rho_{22}\rho_{33}} \ ,
\]
that is, $\rho(\tau)$ is entangled if and only if $c_1(\tau) > 0 $
or $c_2(\tau)>0$. Let us observe that the function $f(\tau)$
controls the evolution of quantum entanglement. Consider for example
$f(\tau) = \varepsilon\gamma e^{-\gamma \tau}$, with $\gamma >0$ and
$\varepsilon \in (0,1]$. One finds from (\ref{alpha-1}) the
following formula $\alpha(\tau) = \varepsilon\gamma[(1-\varepsilon)
e^{\gamma \tau} + \varepsilon]^{-1}$.  Note, that for
$\varepsilon=1$ it reduces to $\alpha(\tau) = \gamma$, that is, it
corresponds to the purely Markovian case. Hence, the parameter
`$1-\varepsilon$' measures the non-Markovianity of the dynamics.
Suppose now that $\rho_0$ is entangled. The entanglement of the
asymptotic state is governed by $\, C(\infty) = 2\max\{
c_1(\infty),c_2(\infty),0\}\,$, with $c_1(\infty) =
(1-\varepsilon)|\rho_{23}| - \sqrt{\rho_{11}\rho_{44}}$ and
$c_2(\infty) = (1-\varepsilon)|\rho_{14}| -
\sqrt{\rho_{22}\rho_{33}}$. It is clear that in the Markovian case
$(\varepsilon=1)$ the asymptotic state is always separable
$(C(\infty)=0)$. However, for sufficiently small `$\varepsilon$'
(i.e. sufficiently big non-Markovianity parameter `$1-\varepsilon$')
one may have $c_1(\infty)>0$ or $c_2(\infty)>0$, that is, the
asymptotic state might be entangled. This example proves the crucial
difference between Markovian and non-Markovian dynamics of composed
systems. In particular controlling `$\varepsilon$' we may avoid
sudden death of entanglement \cite{Eberly}.

 }
\end{example}
\begin{example}  \label{E-PURE}
{\em  Consider the pure decoherence model defined by the following
Hamiltonian  $H=H_R + H_S + H_{SR}$, where $H_R$ is the reservoir
Hamiltonian, $H_S = \sum_n \epsilon_n P_n \; (P_n=|n\>\< n|)$ the
system Hamiltonian and
\begin{equation}\label{}
    H_{SR} = \sum_n P_n \ot B_n
\end{equation}
the interaction part, $B_n=B_n^\dagger$ being reservoirs operators.
The initial product state $\rho \ot \omega_R$ evolves according to
the unitary evolution $e^{-i H t} (\rho \ot \omega_R) e^{i H t}$ and
by partial tracing with respect to the reservoir degrees of freedom
one finds for the evolved system density matrix
\begin{equation*}\label{rho0}
    \rho(t) =
    {\rm Tr}_R [e^{-i H t} (\rho \ot \omega_R) e^{i H t}] =
    \sum_{n,m} c_{mn}(t)P_m\rho P_n \ ,
\end{equation*}
where $c_{mn}(t) = {\rm Tr}( e^{-iZ_m t}\omega_R e^{iZ_n t})$, $Z_n=
\epsilon_n \mathbb{I}_R + H_R + B_n\,$ being reservoir operators.
Note that the matrix $c_{mn}(t)$ is semi-positive definite and hence
\begin{equation}\label{}
    \Lambda(\tau)\,\rho = \sum_{n,m} c_{mn}(\tau) P_m \rho P_n\ .
\end{equation}
 defines the Kraus representation of the completely positive map
$\Lambda(\tau)$. The solution of the pure decoherence model can
therefore be found without explicitly writing down the underlying
master equation. Our method, however, enables one to find the
corresponding generator $\mathcal{L}(\tau)$. It is given by the
following formula
\begin{equation}\label{L-pure}
\mathcal{L}(\tau)\,\rho = \sum_{n,m} \alpha_{mn}(\tau) P_m \rho P_n\
,
\end{equation}
where the functions $\alpha_{mn}(\tau)$ are defined by $\alpha_{mn}=
\dot{c}_{mn}/c_{mn}$. It shows that the pure decoherence model may
be defined by local in time master equation (\ref{local-NM}) with
the non-Markovian generator (\ref{L-pure}). It should be stressed
that this generator is not of the Lindblad form.}
\end{example}

\begin{example}  \label{E-TAN}
{\em Consider the non-Markovian dynamics of a qubit generated by the
following singular generator
\begin{equation}\label{L-3}
\mathcal{L}(\tau) = \tan \tau\, \mathcal{L}_0 \ ,
\end{equation}
with $\mathcal{L}_0$ being the pure dephasing generator defined by
$\mathcal{L}_0\rho = \sigma_z\rho\sigma_z - \rho$. This generator
was analyzed in \cite{Breuer-09,Plenio} in the context of
quantifying non-Markovianity of quantum dynamics. Note that
$\mathcal{L}(\tau)$ has an infinite number of singular points
$\tau_n = (n+\frac 12)\pi$. One easily finds the following perfectly
regular solution for the dynamical map $\Lambda(\tau)= \frac 12
(1+\cos\tau)\oper + \frac 12 (1-\cos\tau)(\mathcal{L}_0+\oper)$,
that is, the density matrix evolves as follows
\begin{equation}\label{EX-3}
\rho(\tau) = \left( \begin{array}{cc} \rho_{11} & \rho_{12}\cos\tau
\\  \rho_{21}\cos\tau & \rho_{22} \end{array}\right) \ ,
\end{equation}
and hence it displays oscillations of the qubit coherence
$\rho_{12}(\tau)$. Note that $\rho(\tau_n)$ is perfectly decohered,
whereas for $\tau = n\pi$ the coherence is perfectly restored.
Finally, one finds extremely simple formula for the corresponding
memory kernel
    $\mathcal{K}(t) = \frac 12\, \mathcal{L}_0$, for $ t\geq t_0$.
Hence, one obtains (\ref{EX-3}) either from the non-local equation
with time-independent memory kernel  $\mathcal{K}(t) = \frac 12\,
\mathcal{L}_0$, or from time-local equation with highly singular
generator (\ref{L-3}).
}
\end{example}
%
%
 In conclusion, we have shown that non-Markovian
quantum evolution may be described either by the non-local equation
(\ref{NZ}) or by a time-local equation (\ref{local-NM}). A similar
strategy based on pseudo-inverse maps have been recently applied in
\cite{Cresser}. We stress, however, that our approach is different.
Clearly, the local approach is technically much simpler, however,
the prize we pay for this simplification is that the corresponding
generator $\mathcal{L}(t-t_0)$ is no longer local in time but it
contains a memory about the starting point `$t_0$'.  Our examples
show that these two descriptions are complementary: if
$\mathcal{K}(\tau)$ is simple (like $\mathcal{K}(t) = \frac 12\,
\mathcal{L}_0$), then $\mathcal{L}(\tau)$ is highly singular (like
in (\ref{L-3})). Vice-versa in the Markovian case
$\mathcal{L}(\tau)= \mathcal{L}_{\rm M}$ but the memory kernel
$\mathcal{K}$ is highly singular and it does involve the Dirac
delta-distribution $\mathcal{K}(\tau) = 2
\delta(\tau)\mathcal{L}_{\rm M}$. Remarkably, singularities of
$\mathcal{L}$ might provide interesting physical content. Note, that
the singularities of `$\tan \tau$' in Example \ref{E-TAN} imply the
interesting features of the dynamical map (\ref{EX-3}): if we evolve
a maximally entangled state $P^+$ of two qubits via the channel
$\Psi(\tau) := \oper \ot \Lambda(\tau)$, then $\Psi(\tau)P^+$ is
separable if and only if $\tau=\tau_n$. It shows that the dynamics
$\Psi(\tau)$ gives rise to entanglement sudden death \cite{Eberly}
whenever $\mathcal{L}(\tau)$ is singular and then entanglement
starts to revive.

This work was partially supported by the Polish Ministry of Science
and Higher Education Grant No 3004/B/H03/2007/33. The authors thank
Jacek Jurkowski for valuable discussion.

\end{document}